\documentclass[12pt]{article}
\usepackage{amstext}
\usepackage{amsmath}
\usepackage{latexsym}
\newcommand{\be}{\begin{equation}}
\newcommand{\ee}{\end{equation}}
\begin{document}
\title{\Large{Path Integral Treatment of Fluctuations in the BCS-BEC Crossover in Superfluid Fermi Gases}
\author{\large E. H. Vivas C\footnote{hvivasc@unal.edu.co}.\\
\it{\normalsize Departamento de F\'{i}sica, Universidad Nacional de Colombia, Sede Manizales}\\
\date{\empty}}}
\maketitle
By using a Path Integral formalism, we study the fluctuating fields in the order parameter for a 3D uniform Fermi gas at $T<Tc$. The BCS-BEC crossover condition is included into the analysis in the framework of the usual $s$-wave scattering approximation. The general expressions for the temperature dependence of the partition function and the quadratic average in the fluctuations fields are explicitly established in terms of the BCS-Gor'kov correlation propagators. 
\\
\\
\centerline{\textbf{\large{Contents}}}
\begin{enumerate}
\item[1.] Abstract \dotfill\ 1
\item[2.] Introduction \dotfill\ 3
\item[3.] The Hubbard-Stratonovich Transformation \dotfill\ 4\\
\hspace{1cm}3.1 Constructing the generating Integral \dotfill\ 5
\item[4.] Gap equation derivation and cut-off criterion \dotfill\ 7\\
\hspace{1cm}4.1 Derivation of the regularization condition in the effective interaction \dotfill\ 9
\item[5.] Expansions in the action functional and TDGL equation \dotfill\ 14
\\
\hspace{1cm}5.1  Static Part \dotfill\ 15\\
\hspace{1cm}5.2  Non-Linear term \dotfill\ 16\\
\hspace{1cm}5.3  Time Dependence of the Linearized Equation \dotfill\ 17
\item[6.] Calculation of the TDGL Coefficients for $T<T_c$ \dotfill\ 20
\\
\hspace{1cm}6.1  Fluctuations in the static approximation \dotfill\ 22\\
\hspace{1cm}6.2 Dynamics of the fluctuations: A generic expression \dotfill\ 24
\\
\hspace{1cm}Concluding Remarks \dotfill\ 25
\\
\hspace{1cm}Appendix A \dotfill\ 26 
\\
\hspace{1cm}Appendix B \dotfill\ 27 
\\
\hspace{1cm}References \dotfill\ 28
\end{enumerate}
\newpage
\centerline{\textbf{\large{2. Introduction}}}
\vspace{0.3cm}
The problem of the crossover from BCS theory with cooperative Cooper pairing to the formation and condensation of composite bosons has attracted considerable attention after the discovery of the high T$_c$ superconductors. Monte Carlo simulations for the attractive (negative $U$) Hubbard model have shown that, in the intermediate coupling, the crossover region displays highly anomalous correlations in degenerate Fermi system above $T_c$ \cite{Randeira0}. More generally, the crossover phenomenon is also of interest for other problems: excitonic condensates, superconductor-insulator transition and Cooper pair formation in resonant Feshbach channels \cite{Griffin1},\cite{Bruun}.
\\
By using the path integral formalism, we study a 3D continuum model of fermions with attractive interactions, at finite temperature. In the first part, we perform a step-by-step derivation of the Time-Dependent Ginzburg Landau (TDGL) equation for the particular case in which the order parameter vanishes ($T=T_c$), and arbitrary strength coupling $g$. We show that in the weak coupling limit, the coefficients in TDGL equation will reduce to the classical ones obtained by Gor'kov (1959) in the framework of the \emph{microscopic} approximation. This formalism is extended for the case $T<T_c$, the $s$-wave pairing symmetry and the Mean-Field Approximation (MFA), following closely the approach of Randeira's group, reported in Ref. \cite{ERS}. 
\\
The main result in this report is the explicit calculation of the quadratic average fluctuation field $\langle\mid\eta\mid^{2}\rangle$, where $\eta$ stands for the fluctuation from the mean field value in the order parameter $\Delta_0$. We obtain the complete set of self-consistent equations for the \emph{dynamical} ($\omega\neq 0$) case for $T<T_c$, and one expression for $\langle\mid\eta\mid^{2}\rangle$, by  keeping only the zero frequency component (the static approximation). Further developments are proposed in terms of numerical analysis, high-order expansion (terms up to fourth order  in the effective action), modification of the \lq\lq starting" Hamiltonian (i.e. by adding the terms \cite{Holland} that describe both the Feshbach resonance effects and preformed Cooper pairs).
\newpage
\centerline{\textbf{\large{3. The Hubbard-Stratonovich Transformation}}}
\vspace{0.5cm}
The Hubbard-Stratonovich (HS) transformation maps certain classes of interacting fermion problem, onto non-interacting electrons  moving through an effective field. Let us consider that the interaction part of the Hamiltonian density involves the product of two bilinear terms:
\begin{equation}
\mathcal{H}_{I}=-g\bar{\Psi}_{\uparrow}\left(x\right)\bar{\Psi}_{\downarrow}\left(x\right)\Psi_{\downarrow}\left(x\right)\Psi_{\uparrow}\left(x\right).
\end{equation}
Taking into consideration the bilinear form for the pair density $A\left(x\right)= \Psi_{\downarrow}\left(x\right)\Psi_{\uparrow}\left(x\right)$,
the Hamiltonian for the interacting part in the continuum space is given by:
\begin{equation}
H_I=\int\mathcal{H}d^{3}\mathbf{x}=\int dx A^{\dag}\left(x\right)A\left(x\right),
\end{equation}
or
\begin{equation}
H_I=-g\sum_{j}A_{j}^{\dag}A_{j}
\end{equation}
on a lattice.\\
The HS transformation replaces the combination $-gA\left(x\right)\rightarrow\Delta\left(x\right)$, in the following way:
\begin{equation}
-gA^{\dag}A\rightarrow A^{\dag}\Delta\left(x\right)+\bar{\Delta}\left(x\right)A\left(x\right)+\frac{\bar{\Delta}\left(x\right)\Delta\left(x\right)}{g}.
\end{equation}
The \emph{field} $\Delta$ can be though as the exchange boson which mediates the original interaction. In this point is necessary to recall several useful identities of complex integration. Bose coherent states are parametrized by complex vectors $\mathbf{z}=\left(z_{1},z_{2},\cdot\cdot\cdot,z_{N}\right)$, $z_{i}=x_{i}+iy_{i}$, where $x_{i},y_{i}$ are real and $N$ is the number of single-particle states. The complex integration is defined as:
\begin{equation}
\int d^{2}\mathbf{z}\equiv\int_{-\infty}^{\infty}\prod_{i}\left(\frac{dx_{i}dy_{i}}{\pi}\right).
\end{equation}
A useful identity of complex integration over a single variable $z$ is:
\begin{equation}
\frac{1}{n!}\int d^{2}z\left(z^{\star}\right)^{n}z^{m}\exp{\left(-z^{\star}z\right)}=\delta_{n,m}.
\end{equation}
Using this definition, it follows that for any complex matrix $G$, whose Hermitian part has only positive eigenvalues, the Gaussian integral is given by:
\begin{equation}
\int d^{2}\mathbf{z}\exp{\left(-\mathbf{z^{\star}}G\mathbf{z}-\mathbf{z^{\star}}_{a}\mathbf{z}-\mathbf{z^{\star}}\mathbf{z}_{b}\right)}=\mbox{det}\mid G\mid^{-1}\exp{\left(\mathbf{z^{\star}}_{a}G^{-1}\mathbf{z}_{b}\right)},
\end{equation}
where $\mathbf{z^{\star}}_{a}$, $\mathbf{z}_{b}$ are any complex vectors. The formalism for fermionic fields is carry out in a similar way (this time the $\mathbf{z}_j$ compounds are Grassman numbers), and Eq. (7) can be unified by the single expression:
\begin{equation}
\int d^{2}\mathbf{z}\exp{\left(-\mathbf{z^{\star}}G\mathbf{z}-\mathbf{z^{\star}}_{a}\mathbf{z}-\mathbf{z^{\star}}\mathbf{z}_{b}\right)}=\mbox{det}\mid G\mid^{-\eta}\exp{\left(\mathbf{z^{\star}}_{a}G^{-1}\mathbf{z}_{b}\right)},
\end{equation}
where $\eta=1$ for bosons and $\eta=-1$ for fermions.
\\
 Under the definitions $\Delta=\Delta_{1}-i\Delta_{2}$, $\bar{\Delta}=\Delta_{1}+i\Delta_{2}$, $d\Delta d\bar{\Delta}\equiv 2id\Delta_{1}d\Delta_{2}$ and with the identities:
\begin{equation}
\int d\Delta_{1}d\Delta_{2}\mbox{\large{e}}^{\left(-\Delta_{1}^2-\Delta_{2}^2\right)/g}=\pi g,\hspace{2cm}\int\frac{d\Delta d\bar{\Delta}}{2\pi ig}\mbox{\large{e}}^{-\bar{\Delta}\Delta/g}=1,
\end{equation}
where $\Delta\left(\mathbf{x},\tau\right)\equiv \Delta\left(x\right)$ is a function of space and time, defined on a grid in space-time that can be made infinitely fine. The natural generalization of Eq. (9) can be written as:
\begin{equation}
\int\mathcal{D}\bigl[\bar{\Delta},\Delta\bigr]\mbox{\large{e}}^{-\int d^{3}\mathbf{x}\int_{0}^{\beta}d\tau\bar{\Delta\left(x\right)}\Delta\left(x\right)/g}=1,
\end{equation}
where $\beta=1/T$ and $\tau$ represents the \emph{imaginary} time in the finite temperature formalism \cite{Fetter}. The integration measurement $\mathcal{D}\bigl[\bar{\Delta},\Delta\bigr]$ is defined as:
\begin{equation}
\mathcal{D}\bigl[\bar{\Delta},\Delta\bigr]\equiv\prod_{\tau,j}\frac{d\bar{\Delta}\left(\mathbf{x}_{j},\tau\right)d\Delta\left(\mathbf{x}_{j},\tau\right)}{\mathcal{N}},
\end{equation}
where $\mathcal{N}=2\pi ig/\left(\Delta^{3}x\Delta\tau\right)$ is a normalization and we have taken the continuum limit $\left(\Delta x,\Delta\tau\right)\rightarrow 0$, replacing the discrete sum over the grid $\sum_{x_{j},\tau_{j}}\bar{\Delta\left(x_j\right)} \Delta\left(x_j\right)$ by the integral $\int d^{3}\mathbf{x}\int_{0}^{\beta}d\tau\bar{\Delta\left(x\right)}\Delta\left(x\right)$.
\\
\\
\centerline{\emph{\textbf{3.1 Constructing the generating integral}}}
\\
\\
The imaginary time generating functional is defined as:
\begin{equation}
\mathcal{Z}=\mbox{Tr}T_{\tau}\left(\exp\Bigl[-\int_{0}^{\beta}d\tau\int d^{3}\mathbf{x}\mathcal{H}\Bigr]\right),
\end{equation}
where Tr$\lbrace\cdot\rbrace$ represents the Trace operator over the quantity $\left(\cdot\right)$, and $\mathcal{H}$ is the Hamiltonian density of the system. By manipulating the Tr and $T_{\tau}$ operators, the partition function can be written as \cite{Auerbach}:
\begin{equation}
\mathcal{Z}=\int\mathcal{D}\bigl[\bar{\Psi},\Psi\bigr]\mbox{\large{e}}^{-S\left(\lbrace\bar{\Psi}\rbrace\lbrace\Psi\rbrace\right)},
\end{equation}
with: 
\begin{equation}
S\left(\lbrace\bar{\Psi}\rbrace\lbrace\Psi\rbrace\right)=\int_{0}^{\beta}\int d^{3}\mathbf{x}\bigl[\bar{\Psi}_{\sigma}\left(x\right)\partial_{\tau}\Psi_{\sigma}\left(x\right)+\mathcal{H}\bigr],
\end{equation}
as the action functional in the fields $\Psi$, $\bar{\Psi}$,  and 
\begin{equation}
\mathcal{H}=\bar{\Psi}_{\sigma}\left(x\right)\Bigl[-\frac{\nabla^{2}}{2m}-\mu\Bigr]\Psi_{\sigma}\left(x\right)+\mathcal{H}_{I},
\end{equation}
where $\mathcal{H}_I$ corresponds to the interacting part for the Hamiltonian density defined in Eq. (1), and $\mu$ is the chemical potential that fixes the average density $n$.
\\
Introducing the \emph{unit} identity in $\mathcal{Z}$:
\begin{equation}
\mathcal{Z}=\int\mathcal{D}\bigl[\bar{\Psi},\Psi\bigr]\int\mathcal{D}\bigl[\bar{\Delta},\Delta\bigr]\mbox{\large{e}}^{-S\left(\lbrace\bar{\Psi}\rbrace\lbrace\Psi\rbrace\right)-\frac{1}{g}\int d^{3}\mathbf{x}\int_{0}^{\beta}d\tau\bar{\Delta}\left(x\right)\Delta\left(x\right)}.
\end{equation}
By inserting the explicit form of $ S\left(\lbrace\bar{\Psi}\rbrace\lbrace\Psi\rbrace\right)$, the exponential argument in Eq. (16) reads:
\begin{eqnarray*}
\hspace{-7cm}S\left(\lbrace\bar{\Psi}\rbrace\lbrace\Psi\rbrace\right)+\frac{1}{g}\int d^{3}\mathbf{x}\int_{0}^{\beta}d\tau\bar{\Delta}\Delta=
\end{eqnarray*}
\begin{eqnarray}
\int d^{3}\mathbf{x}\int_{0}^{\beta}d\tau \Bigl[\bar{\Psi}_{\sigma}\left(\partial_{\tau}-\frac{\nabla^{2}}{2m}-\mu\right)\Psi_{\sigma}-g\bar{\Psi}_{\uparrow}\bar{\Psi}_{\downarrow}\Psi_{\downarrow}\Psi_{\uparrow}+\frac{1}{g}\bar{\Delta}\Delta\Bigr].
\end{eqnarray}
Under the transformations:
\begin{equation}
\Delta\left(x\right)\rightarrow\Delta\left(x\right)+gA\left(x\right);\hspace{1cm} \bar{\Delta}\left(x\right)\rightarrow\bar{\Delta}\left(x\right)+g\bar{A}\left(x\right),
\end{equation}
the last two terms in Eq. (17) become into:
\begin{equation}
-gA^{\dag}A+\frac{1}{g}\bar{\Delta}\Delta\rightarrow \frac{1}{g}\bar{\Delta}\Delta+\bar{\Delta}A+A^{\dag}\Delta=\frac{1}{g}\bar{\Delta}\Delta+\bar{\Delta}\Psi_{\downarrow}\Psi_{\uparrow}+\bar{\Psi}_{\uparrow}\bar{\Psi}_{\downarrow}\Delta.
\end{equation}
This construction must be consistent with the motion equations:   
\begin{eqnarray*}
\left(\partial_{\tau}-\frac{\nabla^{2}}{2m}-\mu\right)\Psi_{\sigma}-g\bar{\Psi}_{\beta}\Psi_{\beta}\Psi_{\sigma}=0,
\end{eqnarray*}
\begin{eqnarray}
\left(-\partial_{\tau}-\frac{\nabla^{2}}{2m}-\mu\right)\bar{\Psi}_{\sigma}-g\bar{\Psi}_{\sigma}\bar{\Psi}_{\beta}\Psi_{\beta}=0,
\end{eqnarray}
derived from the conditions $\delta S/\delta\bar{\Psi}_{\sigma}=0$ and $\delta S/\delta\Psi_{\sigma}=0$ \cite{Landau}.
Then, the l.h.s. of Eq. (17) can be finally written as \footnote{ In this step is important to notice: $\bar{\Psi}_{\sigma}\partial_{\tau}\Psi_{\sigma}=\Psi_{\sigma}\partial_{\tau}\bar{\Psi}_{\sigma}$.}:
\begin{equation}
\int d^{3}\mathbf{x}\int_{0}^{\beta}d\tau\Biggl[\begin{pmatrix}
\bar{\Psi}_{\uparrow}&\Psi_{\downarrow}
\end{pmatrix}
\begin{pmatrix}
-\partial_{\tau}+\frac{\nabla^{2}}{2m}+\mu & \Delta\\
\bar{\Delta} & -\partial_{\tau}-\frac{\nabla^{2}}{2m}-\mu
\end{pmatrix}
\begin{pmatrix}
\Psi_{\uparrow}\\
\bar{\Psi}_{\downarrow}
\end{pmatrix}
+\frac{1}{g}\bar{\Delta}\Delta\Biggr].
\end{equation}
In this procedure we have absorbed the interacting term by replacing it into an effective field which couples the bilinear structure $A$. Integrating out over the fermionic fields, the partition function takes the form: (Cfr. Eqs. (8) and (16))
\begin{equation}
\mathcal{Z}=\int\mathcal{D}\bigl[\bar{\Delta},\Delta\bigr]\mbox{\large{e}}^{-S_{eff}\bigl[\bar{\Delta},\Delta\bigr]},
\end{equation}
where $S_{eff}$ is the effective action which can be written in terms of the (inverse) Nambu propagator $\mathbf{G}^{-1}$ (defined below). The intermediate steps are developed in the following way through the redefinition of Eq. (21):
\begin{equation}
\int d^{3}\mathbf{x}\int_{0}^{\beta}d\tau\bigl[\mathbf{\Psi^{\dag}G^{-1}\Psi}+\frac{1}{g}\bar{\Delta}\Delta\bigr],
\end{equation}
with
\begin{equation}
\mathbf{G^{-1}}=\begin{pmatrix}
-\partial_{\tau}+\frac{\nabla^{2}}{2m}+\mu & \Delta\\
\bar{\Delta} & -\partial_{\tau}-\frac{\nabla^{2}}{2m}-\mu
\end{pmatrix}.
\end{equation}
Recalling Eq. (16):
\begin{equation}
\mathcal{Z}=\int\mathcal{D}\bigl[\bar{\Delta},\Delta\bigr]\Biggl[\mbox{\large{e}}^{-\int d^{3}\mathbf{x}\int_{0}^{\beta}d\tau\frac{1}{g}\bar{\Delta}\Delta}\int\mathcal{D}\bigl[\bar{\Psi},\Psi\bigr]\mbox{\large{e}}^{-\int d^{3}\mathbf{x}\int_{0}^{\beta}d\tau\mathbf{\Psi^{\dag}G^{-1}\Psi}}\Biggr].
\end{equation}
By using Eq. (8) and the expression $\det{\left(\mathbf{M}\right)}$={\large e}$^{\mbox{Tr}\ln{\mathbf{M}}}$, the effective action can be rewritten as: 
\begin{equation}
S_{eff}\bigl[\bar{\Delta},\Delta\bigr]=\int d^{3}\mathbf{x}\int_{0}^{\beta}d\tau\bigl[\frac{1}{g}\bar{\Delta}\Delta-\mbox{Tr}\ln{\mathbf{G^{-1}}\left(\Delta\right)}\bigr].
\end{equation}
\\
\centerline{\large{\textbf{4. Gap Equation Derivation and cut-off criterion}}}
\\
\\
The last integral is analized in the context of static, uniform-saddle points and fluctuations about them. The trivial saddle-point $\Delta=0$, which is stable at sufficiently high T for all couplings, will become unstable below a certain temperature denoted as $T_0$. The transition temperature is defined by the solution of $\delta S_{eff}/\delta\Delta\mid_{\Delta=0}=0$, or, in equivalent way: \cite{Randeira1}, \cite{Randeira2}:
\begin{equation} 
\frac{1}{g}=\sum_{\mathbf{k}}\tanh{\left(\xi_{\mathbf{k}}/2T_{0}\right)}/2\xi_{\mathbf{k}}.
\end{equation}
\emph{Derivation of Eq. (27)}
\\
By applying spatial Fourier transformation in the second term of Eq. (26), we have:
\begin{equation}
S_{eff}\left[\bar{\Delta},\Delta\right]=\int_{0}^{\beta}d\tau\left(\frac{V}{g}\bar{\Delta}\Delta-\ln{\left(\prod_{\mathbf{k}}\det{\left[-\partial_{\tau}-\b{h}_{\mathbf{k},\tau}\right]}\right)}\right),
\end{equation}
and $\b{h}_{\mathbf{k},\tau}$ is a matrix which contains arrays of \emph{kinetic} energy terms $\xi_{\mathbf{k}}=\varepsilon_{\mathbf{k}}-\mu$, with $\varepsilon_{\mathbf{k}}=k^{2}/2m$, as well as the coupling field $\Delta$:
\begin{equation}
\b{h}_{\mathbf{k},\tau}=
\begin{pmatrix}
\xi_{\mathbf{k}}&-\Delta\\
-\bar{\Delta}&-\xi_{\mathbf{k}}
\end{pmatrix}.
\end{equation}
The time-Fourier transformation $\Psi_{\mathbf{k}}\left(\tau\right)=\beta^{-1/2}\sum_{n}\Psi_{\mathbf{k},n}\large{e}^{-i\omega_{n}\tau}$, provides:
\begin{equation}
S_{eff}\left[\bar{\Delta},\Delta\right]\approx\frac{V\beta}{g}\bar{\Delta}\Delta-\ln{\prod_{\mathbf{k},n}\det{\left[i\omega_{n}-\b{h}_{\mathbf{k}}\right]}},
\end{equation}
where the imaginary-time fluctuations dependence in $\Delta$ have been neglected, and $\lbrace\omega_n\rbrace$ represents the set of fermionic Matsubara frequencies defined by: $\omega_{n}=\left(2n+1\right)\pi T$, with $n$ as an integer number. Further manipulations gives: $\lbrace\ln{\prod_{n}\left(\cdot\cdot\cdot\right)}=\sum_{n}\ln\left(\cdot\cdot\cdot\right)$ and $\mid\det\left[i\omega_{n}-\b{h}_{\mathbf{k}}\right]\mid=\omega_{n}^2+\xi_{\mathbf{k}}^2+\Delta^{2}\rbrace$
\begin{eqnarray}
S_{eff}\left[\bar{\Delta},\Delta\right]=\frac{V\beta}{g}\bar{\Delta}\Delta-\sum_{\mathbf{k},n}\ln{\left(\omega_{n}^2+\xi_{\mathbf{k}}^2+\mid\Delta\mid^2\right)}.
\end{eqnarray}
Using the condition $\delta S_{eff}/\delta\bar{\Delta}=0$:
\begin{equation}
\frac{\delta S_{eff}\left[\bar{\Delta},\Delta\right]}{\delta\bar{\Delta}}=\frac{V\beta\Delta}{g}-\sum_{\mathbf{k},n}\frac{\Delta}{\omega_{n}^2+\xi_{\mathbf{k}}^2+\mid\Delta\mid^2}=0,
\end{equation}
or:
\begin{equation}
\frac{1}{g}=\frac{1}{\beta V}\sum_{\mathbf{k},n}\frac{1}{\omega_{n}^2+\xi_{\mathbf{k}}^2+\mid\Delta\mid^2}.
\end{equation}
The complex integration over closed contours give us a simple formula for evaluating sums like (33) \cite{Mattuck}
\begin{equation}
\frac{1}{\beta}\sum_{n}F\left(i\omega_n\right)=\mbox{\texttt{sum of residues of $F\left(\omega\right)f\left(\omega\right)$} at poles of $F\left(\omega\right)$},
\end{equation}
where $f\left(\omega\right)=\left(1+\large{e}^{\beta\omega}\right)^{-1}$ is the Fermi distribution. Clearly, the poles in $F\left(\omega\right)$ are $\pm E_{\mathbf{k}}$, with $E_{\mathbf{k}}=\sqrt{\xi_{\mathbf{k}}^2+\mid\Delta\mid^2}$, therefore the expression (33) can be rewritten as:
\begin{equation}
\frac{1}{g}=-\frac{1}{V}\sum_{\mathbf{k}}\left(\frac{f\left(E_{\mathbf{k}}\right)}{2E_{\mathbf{k}}}-\frac{f\left(-E_{\mathbf{k}}\right)}{2E_{\mathbf{k}}}\right).
\end{equation}
Straightforward calculation leads to:
\begin{equation}
\frac{1}{g}=\frac{1}{V}\sum_{\mathbf{k}}\frac{\tanh{\left(\beta E_{\mathbf{k}}/2\right)}}{2E_{\mathbf{k}}},
\end{equation}
which is reduced to Eq. (27) when $\Delta=0$.
\\
A BSC type cutoff cannot be used to access the strong coupling \emph{Bose regime}. The use of the two-body $s-$wave scattering theory regulates the ultraviolet divergence in the gap equation through the relation:
\begin{equation}
\frac{m}{4\pi a_{s}}=-\frac{1}{g}+\sum_{\mathbf{k}}\frac{1}{2\varepsilon_{\mathbf{k}}},
\end{equation}
where $a_s$ is the scattering length, and $\varepsilon_\mathbf{k}=k^2/2m$. In the following section, we will derive the regularization condition (37) from the elementary $s$-wave scattering theory.
\\
\\
\centerline{\emph{\textbf{4.1 Derivation of the Regularization condition in the Effective Interaction}}}
\\
\\
The concept of the scattering amplitude involves only the directions of the initial and final momenta of the particle undergoing scattering \cite{Landau}. It can therefore be naturally approached by formulating the scattering problem in the momentum representation, where there is no question of the spatial distribution of the process.
\\
Let us transform to the momentum representation the original Schr\"{o}dinger equation
\begin{equation}
-\frac{\hbar^{2}}{2m}\nabla^{2}\psi\left(\mathbf{r}\right)+\left[U\left(\mathbf{r}\right)-E\right]\psi\left(\mathbf{r}\right)=0,
\end{equation}
changing from coordinate to momentum wave functions, i.e. to the Fourier components:
\begin{equation}
a\left(\mathbf{q}\right)=\int\psi\left(\mathbf{r}\right)\mbox{\large{e}}^{-i\mathbf{q\cdot r}}dV.
\end{equation}
Conversely,
\begin{equation}
\psi\left(\mathbf{r}\right)=\int\frac{d^{3}\mathbf{q}}{\left(2\pi\right)^3}a\left(\mathbf{q}\right)\mbox{\large{e}}^{i\mathbf{q\cdot r}}.
\end{equation}
By multiplying (38) by {\large e}$^{-i\mathbf{q\cdot r}}$ and integrate over $dV$. In the first term:
\begin{eqnarray*}
\int dV\mbox{\large{e}}^{-i\mathbf{q\cdot r}}\nabla^{2}\psi\left(\mathbf{r}\right)=\int dV\int\mbox{\large{e}}^{-i\mathbf{q\cdot r}}\times\Biggl[\int\frac{d^{3}\mathbf{q}^{\prime}}{\left(2\pi\right)^3}a\left(\mathbf{q}^{\prime}\right)\left(-q^{\prime}\right)^{2}\mbox{\large{e}}^{i\mathbf{q^{\prime}\cdot r}}\Biggr]
\end{eqnarray*}
\begin{eqnarray}
\hspace{1cm}=\int\frac{d^{3}\mathbf{q}^{\prime}}{\left(2\pi\right)^3}\left(-q^{\prime 2}\right)a\left(\mathbf{q}^{\prime}\right)\delta^{\left(3\right)}\left(\mathbf{q}^{\prime}-\mathbf{q}\right)=-q^{\prime 2}a\left(\mathbf{q}\right).
\end{eqnarray}
The substitution of Eq. (40) into the second term of Eq. (38)\footnote[1]{Here, we must insert $U\left(\mathbf{r}\right)=\int\frac{d^3\mathbf{k}}{\left(2\pi\right)^3}U\left(\mathbf{k}\right)\mbox{\large{e}}^{i\mathbf{k\cdot r}}$, and $\left(2\pi\right)^3\delta^{\left(3\right)}\left(\mathbf{k+q^{\prime}-q}\right)=\int dV \mbox{\large{e}}^{i\left(\mathbf{k+q^{\prime}-q}\right)}.$} gives:
\begin{equation}
\int U\left(\mathbf{r}\right)\psi\left(\mathbf{r}\right)\mbox{\large{e}}^{-i\mathbf{q^{\prime}\cdot r}}dV=\int\frac{d^{3}\mathbf{q}^{\prime}}{\left(2\pi\right)^3}
U\left(\mathbf{q}-\mathbf{q}^{\prime}\right)a\left(\mathbf{q}^{\prime}\right).
\end{equation}
Thus Schr\"{o}dinger's equation in the momentum representation becomes:
\begin{equation}
\left(\frac{\hbar^{2}q^{2}}{2m}-E\right)a\left(\mathbf{q}\right)+\int\frac{d^{3}\mathbf{q}^{\prime}}{\left(2\pi\right)^3}
U\left(\mathbf{q}-\mathbf{q}^{\prime}\right)a\left(\mathbf{q}^{\prime}\right)=0.
\end{equation}
The wave function describing the scattering of particles with momentum $\hbar\mathbf{k}$ has the form:
\begin{equation}
\psi_{\mathbf{k}}\left(\mathbf{r}\right)=\mbox{\large{e}}^{i\mathbf{k\cdot r}}+\chi_{\mathbf{k}}\left(\mathbf{r}\right),
\end{equation}
where $\chi_{\mathbf{k}}\left(\mathbf{r}\right)$ is a function whose asymptotic form (as $r\rightarrow \infty$) is that of an outgoing spherical wave. Its Fourier component is: (see Eq. 39)
\begin{equation}
a_{\mathbf{k}}\left(\mathbf{q}\right)=\delta^{\left(3\right)}\left(\mathbf{q-k}\right)+\chi_{\mathbf{k}}\left(\mathbf{q}\right)
\end{equation}
and substitution in (43) gives the following equation for the function $\chi_{\mathbf{k}}\left(\mathbf{q}\right)$:\\($\left(q^{2}-k^{2}\right)\delta^{3}\left(\mathbf{q-k}\right)=0$)
\begin{equation}
\frac{\hbar^2}{2m}\left(k^2-q^2\right)\chi_{\mathbf{k}}\left(\mathbf{q}\right)=U\left(\mathbf{q-k}\right)+\int\frac{d^{3}q^{\prime}}{\left(2\pi\right)^3}U\left(\mathbf{q-q^{\prime}}\right)\chi_{\mathbf{k}}\left(\mathbf{q}^{\prime}\right),
\end{equation}
where $E=\hbar^{2}k^{2}/2m$.
This equation may conveniently be transformed by using instead of $\chi_{\mathbf{k}}\left(\mathbf{q}\right)$ another unknown function defined by:
\begin{equation}
\chi_{\mathbf{k}}\left(\mathbf{q}\right)=\frac{2m}{\hbar^{2}}\frac{F\left(\mathbf{k,q}\right)}{q^2-k^2-i0}.
\end{equation}
This eliminates the singularity at $q^2=k^2$ in the coefficients of (46), which becomes:
\begin{equation}
F\left(\mathbf{k,q}\right)=-U\left(\mathbf{q-k}\right)-\frac{2m}{\hbar^{2}}\int\frac{d^{3}q^{\prime}}{\left(2\pi\right)^3}\frac{U\left(\mathbf{q-q^{\prime}}\right)F\left(\mathbf{k,q^{\prime}}\right)}{q^{\prime 2}-k^2-i0}.
\end{equation}
The term $i0$, which denotes the limit of $i\delta$ as $\delta\rightarrow +0$, is included in the (47) definition in order to give a definite sense to the integral (48), since it establishes the manner of passage round the pole $q^{\prime 2}=k^2$. The Fourier transformation gives:
\begin{equation}
\chi_{\mathbf{k}}\left(\mathbf{r}\right)=\frac{2m}{\hbar^{2}}\int\frac{d^{3}q}{\left(2\pi\right)^3}
\frac{F\left(\mathbf{k,q}\right)\mbox{\large{e}}^{i\mathbf{q\cdot r}}}{q^2-k^2-i0}.
\end{equation}
Using $d^{3}q=q^{2}dq d\Omega_{\mathbf{q}}$, and first integrate over $d\Omega_{\mathbf{q}}$ i.e., over the directions of the vector $\mathbf{q}$ relative to $\mathbf{r}$:
\begin{equation}
\chi_{\mathbf{k}}\left(\mathbf{r}\right)=-\frac{im}{2\pi^{2}\hbar^{2}r}\int_{-\infty}^{\infty}\frac{F\left(\mathbf{k},q\mathbf{n^{\prime}}\right)\mbox{\large{e}}^{iqr}qdq}{q^2-k^2-i0},
\end{equation}
where $\mathbf{n^{\prime}}=\mathbf{r}/r$.
The integrand has poles at points $q=k+i0$ and $q=-k-i0$; the path of integration in the complex $q-$plane passes respectively below and above these values. The integral tends to zero as $r\rightarrow \infty$, and the integral round the loop is given by $2\pi i$ times the residue of the integrand at the pole $q=k$. The final result is:
\begin{equation}
\chi_{\mathbf{k}}\left(\mathbf{r}\right)=\frac{m}{2\pi\hbar^{2}}\frac{\mbox{\large{e}}^{ikr}}{r}F\left(k\mathbf{n},k\mathbf{n^{\prime}}\right),
\end{equation}
where $\mathbf{n}$ is a unit vector in the direction of $\mathbf{k}$. We have derived the requiered asymptotic form of the wave function, and the scattering amplitude is:
\begin{equation}
f\left(\mathbf{n,n^{\prime}}\right)=\frac{m}{2\pi\hbar^{2}}F\left(k\mathbf{n},k\mathbf{n^{\prime}}\right).
\end{equation}
In the first approximation, omitting the integral term in (48), we have $F\left(\mathbf{k,q}\right)=-U\left(\mathbf{q-k}\right)$. In the next approximation, we substitute it in the integral equation, then:
\begin{equation}
f\left(\mathbf{n,n^{\prime}}\right)=-\frac{m}{2\pi\hbar^{2}}\Biggl[U\left(\mathbf{k-k^{\prime}}\right)+\frac{2m}{\hbar^2}\int\frac{d^{3}k^{\prime\prime}}{\left(2\pi\right)^3}\frac{U\left(\mathbf{k^{\prime}}-\mathbf{k^{\prime\prime}}\right)U\left(\mathbf{k^{\prime\prime}}-\mathbf{k}\right)}{k^2-k^{\prime\prime 2}+i0}\Biggr],
\end{equation}
with $\mathbf{k}=k\mathbf{n}$, $\mathbf{k}^{\prime}=k\mathbf{n^{\prime}}$. The first term is often called \emph{the Born approximation}. In the second approximation, the scattering amplitude does not have the symmetry property $f\left(\mathbf{k,k^{\prime}}\right)=f^{\star}\left(\mathbf{k^{\prime},k}\right)$. At first sight it may seem that the integral term in (53) is also symmetrical with respect to interchange of the initial and final states. However, this symmetry does not exist, because the path of integration and the direction of the passage round the pole are altered when the complex conjugate expression is taken.
\\
In order to make these ideas quantitative, consider the the two-particles scattering, where the particles are assume to have equal masses $m$, and therefore $m_{r}=m/2$. In this case, the scattering matrix $F\left(\mathbf{k,q}\right)$ in Eq. (48) satisfies the so called Lippmann-Schwinger equation \cite{PS}:
\begin{equation}
F\left(\mathbf{k,q}\right)=-U\left(\mathbf{q-k}\right)-\int\frac{d^{3}q^{\prime}}{\left(2\pi\right)^3}\frac{U\left(\mathbf{q-q^{\prime}}\right)F\left(\mathbf{k,q^{\prime}}\right)}{\hbar q^{\prime 2}/m-E-i0},
\end{equation}
where $E=\hbar^2k^2/m$ is the energy eigenvalue and $U\left(\mathbf{q-k}\right)$ is the Fourier transform of the bare atom-atom interaction. The scattered wave at large distances and for zero energy  ($E=0, k=0$), may be calculated from (54). In this case, the previous formulation must be slightly modified through the change $2m\rightarrow m$ \cite{Abrikosov}. Using the formula:
\begin{equation}
\int\frac{d^{3}k}{\left(2\pi\right)^3}
\frac{\mbox{\large{e}}^{i\mathbf{k\cdot\left(r-r^{\prime}\right)}}}{k^2}=\frac{1}{4\pi\mid\mathbf{r-r^{\prime}}\mid},
\end{equation}
expression (49) can be approximate by:
\begin{equation}
\chi_{0}\left(\mathbf{r}\right)\approx-\frac{mU\left(0\right)}{4\pi\hbar^2 r}\equiv-\frac{a_s}{r}.
\end{equation}
In the context of Born approximation, which is obtained by taking only the first term on the right side of (54), the scattering length is given by
\begin{equation}
a_{s}=\frac{mU\left(0\right)}{4\pi\hbar^{2}}.
\end{equation}
The relation (57) between $U$ and the scattering amplitude is not exact, but is only valid up to first-order terms. We are also interested in higher order terms in the energy, the formula (57) has to be corrected.  Using Eq. (54): ($F\approx-4\pi\hbar a_{s}/m$) in the l.h.s, and $F\approx -U\left(0\right)$ in r.h.s)
\begin{equation}
-\frac{4\pi\hbar^{2}a_{s}}{m}\approx-U\left(0\right)+\frac{U^{2}\left(0\right)}{V}\sum_{\mathbf{q}}\frac{1}{\hbar q^{2}/m}.
\end{equation}           
Further arrangement finally leads to the expression (37):             
\begin{equation}
\frac{m}{4\pi\hbar^{2}\mid a_{s}\mid}\approx -\frac{1}{U\left(0\right)}+\frac{1}{V}\sum_{\mathbf{q}}\frac{1}{\hbar q^{2}/m}.
\end{equation}
Eliminating the coupling constant $g\equiv U\left(0\right)$ in Eqs. (27) and (59), we obtain:
\begin{equation}
-\frac{m}{4\pi\hbar^{2}\mid a_{s}\mid}=\sum_{\mathbf{q}}\left(\frac{\tanh{\left(\xi_{\mathbf{q}}/2T_{0}\right)}}{2\xi_{\mathbf{q}}}-\frac{1}{2\varepsilon_{\mathbf{q}}}\right).
\end{equation}
In order to solve the last equation for $T_0$, we must determine the chemical potential $\mu$ as a function of the coupling interaction $g$ and the temperature, by using the relationship $N=-\partial\Omega/\partial\mu$. The saddle point approximation for the thermodynamic potential, $\Omega_{0}=S_{eff}\left[\Delta\right]/\beta$, leads to the number equation: (see Eq. (31))
\begin{equation}
n_{\mbox{\tiny{fermions}}}\left(T,\mu\right)=-\frac{\partial\Omega}{\partial\mu}=-\frac{1}{\beta}\sum_{\mathbf{k},n}\frac{2\xi_{\mathbf{k}}}{\omega_{n}^{2}+\xi_{\mathbf{k}}^{2}+\mid\Delta\mid^{2}}=\sum_{\mathbf{k}}\frac{\xi_{\mathbf{k}}}{E_{\mathbf{k}}}\tanh{\left(\frac{\beta E_{\mathbf{k}}}{2}\right)},
\end{equation}
with $E_{\mathbf{k}}=\sqrt{\xi_{\mathbf{k}}^{2}+\Delta^{2}}$.
For bosons:
\begin{equation}
n\left(T,\mu\right)=\sum_{\mathbf{k}}\Bigl[1-\frac{\xi_{\mathbf{k}}}{E_{\mathbf{k}}}\tanh{\left(\frac{\beta E_{\mathbf{k}}}{2}\right)}\Bigr].
\end{equation}
\newpage
\centerline{\large{\textbf{5. Expansions in the Action Functional and TDGL Equation}}}
\vspace{0.5cm}
In order to obtain the contributions on the fourth-order fluctuations in the effective action $S_{eff}\left[\bar{\Delta},\Delta\right]$ due to the bosonic fields $\Delta\left(\bar{\Delta}\right)$, we must start by writting the Nambu propagator (24) as: \cite{Randeira3}
\begin{equation}
\mathbf{G^{-1}}=-\partial_{\tau}-\mathcal{K}\sigma_{z}+\Delta\sigma^{+}+\bar{\Delta}\sigma^{-}=\mathbf{G_0}^{-1}+\Delta\sigma^{+}+\bar{\Delta}\sigma^{-},
\end{equation}
where $\mathcal{K}=-\nabla^{2}/2m-\mu$, $\sigma^{\pm}=\left(\sigma_{x}\pm i\sigma_{y}\right)/2$, where $\sigma_{j}$'s are the Pauli's matrices, and $\mathbf{G_0}^{-1}$ is the \emph{non-interacting} propagator. The expression for $S_{eff}$ in Eq. (26) becomes:
\begin{equation}
S_{eff}\left[\bar{\Delta},\Delta\right]=\int d^{3}\mathbf{x}\int_{0}^{\beta}d\tau\Biggl[\frac{\mid\Delta\mid^2}{g}-\mbox{Tr}\ln{\mathbf{G_0}^{-1}}-\mbox{Tr}\ln{\left[1+\mathbf{G_{0}}\left(\Delta\sigma^{+}+\bar{\Delta}\sigma^{-}\right)\right]}\Biggr].
\end{equation}
Using the identity $\mbox{Tr}\ln{\left(1+A\right)}=-\sum_{n=1}^{\infty}\left(-1\right)^{n}\mbox{Tr}\left(A\right)^{n}/n$, we obtain the combination:
\begin{equation}
\mbox{Tr}\ln{\left[1+\mathbf{G_{0}}\left(\Delta\sigma^{+}+\bar{\Delta}\sigma^{-}\right)\right]}\rightarrow-\frac{1}{2}\mbox{Tr}\left(\mathbf{G_{0}}\sigma^{+}\mathbf{G_{0}^{\prime}}\sigma^{-}\Delta\bar{\Delta}^{\prime}+\mathbf{G_{0}}\sigma^{-}\mathbf{G_{0}^{\prime}}\sigma^{+}\Delta^{\prime}\bar{\Delta}\right),
\end{equation}
for the second order expansion in the last term of (64) ($\sigma^{\pm}\mathbf{G_0}\sigma^{\pm}=0$).
Rewritting (64) in terms of the Fourier variables $\left(i\omega_{m},i\nu_{n},\mathbf{k,q}\right)$, with $i\omega_m=2\pi T\left(m+1/2\right)$, $\left(i\nu_{n}=2\pi Tn\right)$ as the Fermionic (Bosonic) Matsubara frequencies, the Green's function reads:
\begin{equation}
\mathbf{G_0}=
\begin{pmatrix}
\left(i\omega_{m}-\xi_{\mathbf{k}}\right)^{-1}&0\\
0&\left(i\omega_{m}+\xi_{\mathbf{k}}\right)^{-1}
\end{pmatrix},
\end{equation}
and the effective action becomes into:
\begin{equation}
S_{eff}\left[\bar{\Delta},\Delta\right]=S_0+\sum_{q}\Biggl[\frac{\mid\Delta\left(q\right)\mid^2}{g}+\frac{1}{2}\mbox{Tr}\left(\mathbf{G_{0}}_{k}\sigma^{+}\mathbf{G_{0}}_{-k+q}\sigma^{-}+\mathbf{G_{0}}_{-k+q}\sigma^{-}\mathbf{G_{0}}_{k}\sigma^{+}\right)\mid\Delta\left(q\right)\mid^2\Biggr]
\end{equation}
with $S_{0}=-\int d^{3}\mathbf{x}\int_{0}^{\beta}d\tau\mbox{Tr}\ln{\mathbf{G_0}^{-1}}$, and the Tr operator involves the \emph{internal}-vertex sumation over the fermionic four-momenta $k$. The operation $\frac{1}{2}$ Tr $\left(\cdot\cdot\cdot\right)$ finally gives:
\begin{equation}
\frac{1}{2}\mbox{Tr}\left(\cdot\cdot\cdot\right)=\sum_{k}\left(i\omega_{m}-\xi_{\mathbf{k}}\right)^{-1}\left(i\omega_{m}-i\nu_{n}+\xi_{\mathbf{-k+q}}\right)^{-1}.
\end{equation}
Expression (67) can be reduced to:
\begin{equation}
S_{eff}\left[\bar{\Delta},\Delta\right]=S_{0}+\sum_{q}\frac{\mid\Delta\left(q\right)\mid^2}{\Gamma\left(q\right)},
\end{equation}
for the Gaussian approximation. The coefficient $\Gamma_{q}^{-1}$ is defined by:
\begin{equation}
\Gamma_{q}^{-1}=-\frac{m}{4\pi a_s}+\sum_{k}\Bigl[\left(i\omega_{m}-\xi_{\mathbf{k}}\right)^{-1}\left(i\omega_{m}-i\nu_{n}+\xi_{\mathbf{-k+q}}\right)^{-1}+\frac{1}{2\varepsilon_{\mathbf{k}}}\Bigr],
\end{equation}
where regularization condition for the effective $s$-wave interaction $g$ (See Eq. 37) has been inserted in the last expression. The resulting equation for the thermodynamic potential $\Omega=\Omega_{0}-\beta^{-1}\sum_{q}\ln{\Gamma_{q}^{-1}}$ is identical to the diagrammatic result of Nozi\`{e}res and Schmitt-Rink (NSR), in the framework of particle-particle scattering $t$-matrix theory \cite{NSR}.
\\
\\
\centerline{\textbf{\emph{5.1 Static Part}}}
\\
\\
Expansions for the static compound of $\Gamma^{-1}\left(\mathbf{q},0\right)$ can be written as: $\Gamma^{-1}\left(\mathbf{q},0\right)\approx a+\\ c\mid\mathbf{q}\mid^{2}/2m+\cdot\cdot\cdot$, where $a$  is evidently given by:
\begin{eqnarray*}
a=\Gamma^{-1}\left(\mathbf{0},0\right)=-\frac{m}{4\pi a_{s}}+\sum_{k}\Bigl[\left(i\omega_{m}-\xi_{\mathbf{k}}\right)^{-1}\left(i\omega_{m}+\xi_{\mathbf{-k}}\right)^{-1}+\frac{1}{2\varepsilon_{\mathbf{k}}}\Bigr]
\end{eqnarray*}
\begin{eqnarray}
=-\frac{m}{4\pi a_s}+\sum_{\mathbf{k}}\Bigl[\frac{1}{2\varepsilon_{\mathbf{k}}}-\frac{X}{2\xi_{\mathbf{k}}}\Bigr], \hspace{2cm} X=\tanh{\left(\frac{\beta\xi_{\mathbf{k}}}{2}\right)}.
\end{eqnarray}
The calculation for the $c$-coefficient starts from Eq. (70):
\begin{equation}
\Gamma^{-1}\left(\mathbf{q},0\right)=-\frac{m}{4\pi a_{s}}+\sum_{\mathbf{k}}\Bigl[\frac{1}{2\varepsilon_{\mathbf{k}}}-\frac{1-n_{F}\left(\xi_{\mathbf{k}}\right)-n_{F}\left(\xi_{\mathbf{-k+q}}\right)}{\xi_{\mathbf{k}}+\xi_{\mathbf{-k+q}}}\Bigr].
\end{equation}
Using the relations:
\begin{eqnarray*} 
1-n_{F}\left(\xi_{\mathbf{k}}\right)-n_{F}\left(\xi_{\mathbf{-k+q}}\right)=\frac{1}{2}\Bigl[\tanh{\left(\frac{\beta\xi_{\mathbf{k}}}{2}\right)}+\tanh{\frac{\beta}{2}\left(\xi_{\mathbf{k}}+h\right)}\Bigr],
\end{eqnarray*}
\begin{eqnarray*}
h=\frac{1}{2m}\left(q^2-2q\left(\mathbf{k\cdot\widehat{n}}\right)\right),
\end{eqnarray*}
valid in the small momentum $q$ approach, Eq. (72) becomes into: 
\begin{equation}
1-n_{F}\left(\xi_{\mathbf{k}}\right)-n_{F}\left(\xi_{\mathbf{-k+q}}\right)\approx \frac{1}{2}\left[2X+\frac{1}{2}\beta hY-\left(\frac{1}{2}\beta h\right)^{2}XY\right],
\end{equation}
where $Y=\mbox{sech}^2{\left(\beta\xi_{\mathbf{k}}/2\right)}$.
It is also useful to consider:
\begin{equation}
\frac{1}{\xi_{\mathbf{k}}+\xi_{\mathbf{-k+q}}}\approx\frac{1}{2\xi_{\mathbf{k}}}\left(1-\frac{h}{2\xi_{\mathbf{k}}}+\frac{h^2}{4\xi_{\mathbf{k}}^2}\right).
\end{equation}
Straightforward algebra leads to the $\mathcal{O}\left(q^2\right)$ coefficient in $\Gamma^{-1}\left(\mathbf{q},0\right)$:
\begin{equation}
c\frac{q^2}{2m}\equiv\Biggl[\frac{X}{4\xi_{\mathbf{k}}^2}-\frac{\left(\mathbf{k\cdot\widehat{n}}\right)^{2}X}{4m\xi_{\mathbf{k}}^3}-\frac{\beta Y}{8\xi_{\mathbf{k}}}+\frac{\beta\left(\mathbf{k\cdot\widehat{n}}\right)^{2}Y}{8m\xi_{\mathbf{k}}^2}+\frac{\beta^{2}\left(\mathbf{k\cdot\widehat{n}}\right)^{2}XY}{8m\xi_{\mathbf{k}}}\Biggr]\left(\frac{q^2}{2m}\right).
\end{equation}
\centerline{\emph{\textbf{5.2 Non-Linear term}}}
\\
\\
The fourth order fluctuations in $\mathcal{O}\left(\Delta^4\right)$ for the effective action $S_{eff}$ can be calculated in analogous way. In this case, we must analyze the non-zero contributions:
\begin{eqnarray*}
\mathbf{G_0}\sigma^{+}\mathbf{G_{0}^{\prime}}\sigma^{-}\mathbf{G_{0}^{\prime\prime}}\sigma^{+}\mathbf{G_{0}^{\prime\prime\prime\prime}}\sigma^{-}+ \mbox{\texttt{permutations}}
\end{eqnarray*}
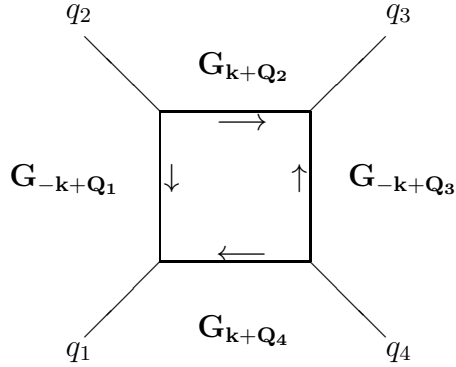
\setlength{\unitlength}{1cm}
\begin{picture}(7.5,7.5)
\put(4,4){$\mathbf{G}_{\mathbf{-k+Q_{1}}}$}
\put(8.5,4){$\mathbf{G}_{\mathbf{-k+Q_{3}}}$}
\put(6.5,5.5){$\mathbf{G}_{\mathbf{k+Q_{2}}}$}
\put(6.5,2.0){$\mathbf{G}_{\mathbf{k+Q_{4}}}$}
\put(6,5){\line(-1,1){1}}
\put(6.75,4.75){$\longrightarrow$}
\put(6.75,3){$\longleftarrow$}
\put(6.05,4){$\downarrow$}
\put(7.75,4){$\uparrow$}
\put(4.75,6.25){$q_2$}
\put(8,5){\line(1,1){1}}
\put(9,6.25){$q_3$}
\put(8,3){\line(1,-1){1}}
\put(9,1.75){$q_4$}
\put(6,3){\line(-1,-1){1}}
\put(4.75,1.75){$q_1$}
\put(6,5){\line(1,0){2}}
\put(8,5){\line(0,-1){2}}
\put(8,3){\line(-1,0){2}}
\put(6,5){\line(0,-1){2}}
\end{picture}
\\
In the our case, we have interest in the $q_{1}=q_{2}=q_{3}=0$ coefficient: $b\left(0,0,0\right)$ which is defined by:
\begin{eqnarray*}
-\frac{1}{4}\mbox{Tr}\left(\mathbf{G_{0}}_{k}\sigma^{+}\mathbf{G_{0}}_{-k}\sigma^{-}\mathbf{G_{0}}_{k}\sigma^{+}\mathbf{G_{0}}_{-k}\sigma^{-}+\mathbf{G_{0}}_{-k}\sigma^{-}\mathbf{G_{0}}_{k}\sigma^{+}\mathbf{G_{0}}_{-k}\sigma^{-}\mathbf{G_{0}}_{k}\sigma^{+}\right)=
\end{eqnarray*}
\begin{eqnarray}
=-\frac{1}{2}\sum_{k}
\left(i\omega_{m}-\xi_{\mathbf{k}}\right)^{-2}\left(i\omega_{m}+\xi_{\mathbf{k}}\right)^{-2}=-\frac{1}{2}b\left(0,0,0\right).
\end{eqnarray}
The expression (69) must be formally rewritten as:
\begin{equation}
S_{eff}\left[\bar{\Delta},\Delta\right]=S_{0}+\sum_{q}\frac{\mid\Delta\left(q\right)\mid^2}{\Gamma\left(q\right)}+\frac{1}{2}\sum_{q_{1},q_{2},q_{3}}b\left(0,0,0\right)\Delta\left(q_{1}\right)\bar{\Delta}\left(q_{2}\right)\Delta\left(q_{3}\right)\bar{\Delta}\left(q_{1}+q_{3}-q_{2}\right).
\end{equation}
with
\begin{equation}
b\equiv b\left(0,0,0\right)=\sum_{\mathbf{k}}\frac{X}{4\xi_{\mathbf{k}}^3}.
\end{equation}
To study the evolution  of the TDGL we recall the relationship $\delta S_{eff}/\delta\Delta^{\star}=0$ near to $T_c$ and for $\Delta$ slowly varying in the space and time. Finally, the effective action reads:
\begin{equation}
S_{eff}\left[\bar{\Delta},\Delta\right]\approx S_{0}+\left(a+c\frac{q^2}{2m}\right)\mid\Delta\mid^{2}+\frac{b}{2}\mid\Delta\mid^{4}.
\end{equation}
\centerline{\emph{\textbf{5.3 Time Dependence of the Linearized Equation}}}
\\
\\
The linearized dependence requieres a careful examination of Gaussian fluctuations about the broken symmetry solution, and a simple low frequency expansion is obtained when $\Delta\left(T\right)<<\omega$. In this case, there is necessary to expand $Q\left(iq_{l}\right)=\Gamma^{-1}\left(\mathbf{q}=0,iq_{l}\right)-\Gamma^{-1}\left(\mathbf{q}=0,0\right)$ in powers of $\omega$ after analytic continuation. From Eq. (70), we have:
\begin{equation}
Q\left(iq_l\right)=\sum_{\mathbf{k}}\tanh{\left(\frac{\beta\xi_{\mathbf{k}}}{2}\right)}\Bigl[\frac{1}{2\xi_{\mathbf{k}}-iq_l}-\frac{1}{2\xi_{\mathbf{k}}}\Bigr].
\end{equation}
Under $iq_{l}\rightarrow\omega+i\epsilon$, and by using the relation $\left(x\pm i\epsilon\right)^{-1}=\mathcal{P}\left(\frac{1}{x}\right)\mp i\pi\delta\left(x\right)$, with $\mathcal{P}$ as the principal value, the frequency-dependent component becomes:
\begin{equation}
Q\left(\omega\right)=\mathcal{P}\Biggl[\sum_{\mathbf{k}}\tanh{\left(\frac{\beta\xi_{\mathbf{k}}}{2}\right)}\frac{\omega}{2\xi_{\mathbf{k}}\left(2\xi_{\mathbf{k}}-\omega\right)}\Biggr]+i\pi\sum_{\mathbf{k}}\tanh{\left(\frac{\beta\xi_{\mathbf{k}}}{2}\right)}\delta\left(2\xi_{\mathbf{k}}-\omega\right).
\end{equation}
The operator sum can be transformed into the integral through $\sum_{\mathbf{k}}\left(\cdot\cdot\cdot\right)=\\ \int d\varepsilon N\left(\varepsilon_{F}\right)\varepsilon_{F}^{-1/2}\varepsilon^{1/2}\left(\cdot\cdot\cdot\right)$. Straightforward calculation leads to: $Q\left(\omega\right)=Q^{\prime}\left(\omega\right)+iQ^{\prime\prime}\left(\omega\right)$, where:
\begin{eqnarray*}
Q^{\prime}\left(\omega\right)=
\frac{N\left(\varepsilon_{F}\right)}{\varepsilon_{F}}\mathcal{P}\int_{0}^{\infty}
d\varepsilon\frac{\omega\varepsilon^{1/2}\tanh{\left(\frac{\varepsilon-\mu}{2T_c}\right)}}{2\left(\varepsilon-\mu\right)\left(2\varepsilon-2\mu-\omega\right)}, 
\end{eqnarray*}
\begin{eqnarray}
Q^{\prime\prime}\left(\omega\right)=\frac{\pi N\left(\varepsilon_{F}\right)}{\sqrt{\varepsilon_{F}}}\tanh{\left(\frac{\omega}{4T_c}\right)}\frac{\sqrt{\omega+2\mu}}{2^{3/2}}\Theta\left(\omega+2\mu\right).
\end{eqnarray}
In the low frequency limit $\left(\omega<<T_c\right)$, an expansion  of  $Q$ is possible for both the BCS and BEC regimes, where the condition $\omega<<\mid\mu\mid$ is satisfied. Linearization of $Q$  finally leads to:
\begin{equation}
Q\left(\omega\right)=d\omega,
\end{equation}
with
\begin{equation}
d=\sum_{\mathbf{k}}\frac{\tanh{\left(\beta_c\xi_{\mathbf{k}}/2\right)}}{4\xi_{\mathbf{k}}^2}+i\frac{\pi\beta_{c}N\left(\varepsilon_{F}\right)\sqrt{\mu}}{8\varepsilon^{1/2}_{F}}\Theta\left(2\mu\right).
\end{equation}
The inverse Fourier transformation in the variables $\left(\mathbf{x},t\right)$ allows to write the TDGL equation: ($q^2\rightarrow -\nabla^2$, $\omega\rightarrow -i\partial/\partial t$)
\begin{equation}
\left(a-c\frac{\nabla^2}{2m}+b\mid\Delta\mid^2-id\frac{\partial}{\partial t}\right)\Delta\left(\mathbf{x},t\right)=0.
\end{equation}
For weak coupling approximation, the cut off frequency  is much larger than $T_c$ but considerably smaller than $\varepsilon_F$. The energy cut off is equivalent to truncating the Matsubara sums at a limiting number $N_0=\Omega_{BCS}/2\pi T>>1$ \cite{Kopnin}. The parameter $a$ can be therefore rewritten as (for excitations near to the Fermi energy $\varepsilon_F$):
\begin{equation}
\bar{\Gamma}^{-1}\left(0,0\right)=\frac{2}{\beta}\sum_{n>0}\int d\varepsilon\frac{N\left(\varepsilon_F\right)\sqrt{\varepsilon}}{\sqrt{\varepsilon_F}}\frac{1}{\left(i\omega_n+\varepsilon\right)}\frac{1}{\left(i\omega_n-\varepsilon\right)}\approx\frac{2\pi N\left(\varepsilon_F\right)}{\beta}\sum_{n>0}\frac{1}{\omega_n}.
\end{equation}
The order parameter vanishes at the critical temperature $T=T_c$, therefore the following condition holds:
\begin{equation}
\frac{1}{g}=2\pi T\sum_{n=0}^{N_0}\frac{1}{\sqrt{\omega_{n}^2+\mid\Delta\mid^2}}\mid_{\Delta=0}\approx\ln{\left(\frac{\Omega_{BCS}}{2\pi T_c}\right)}+\ln{4}+\gamma,
\end{equation}
where $\Omega_{BCS}$ is, in general, the characteristic cut off frequency in the superfluid state. Then
\begin{equation}
2\pi T\sum_{n=0}^{N_{0}\left(T\right)}\frac{1}{\omega_n}=\frac{1}{g}+\sum_{N_{0}\left(T_c\right)}^{N_{0}\left(T\right)}\frac{1}{n+1/2}=\frac{1}{g}+\ln{\frac{T_c}{T}}.
\end{equation}
 The combination of Equations (71) and (88) leads to: 
\begin{equation} 
a\equiv N\left(\varepsilon_F\right)\ln{\frac{T}{T_c}}.
\end{equation}
For the non-linear coefficient $b$, we recall the expression (76) as:
\begin{eqnarray*}
b=\frac{2}{\beta}\sum_{n\geq 0}N\left(\varepsilon_{F}\right)\int d\varepsilon\frac{1}{\left(\omega_{n}^2+\varepsilon^{2}\right)^2}.
\end{eqnarray*}
After integration over the upper-half complex plane in the variable $\varepsilon$, we find:
\begin{eqnarray*}
b\approx \frac{1}{2}N\left(\varepsilon_{F}\right)\Biggl[\frac{1}{4\pi^{2}T_{c}^{2}}\sum_{n\geq 0}\frac{1}{\left(n+1/2\right)^3}\Biggr].
\end{eqnarray*}
Using $\sum_{0}^{\infty}\left(n+1/2\right)^{-z}=\left(2^{z}-1\right)\zeta\left(z\right)$, with $\zeta\left(z\right)$ as the Riemmann-Zeta function, finally we obtain the quartic contribution term on TDGL: $b=7N\left(\varepsilon_F\right)\zeta\left(3\right)/8\pi^{2}T_{c}^2$. In analogous way, for weak coupling interaction, the parameter $c$ can be obtained from Eq. (70). The term which contains $q$ is defined through the expansion of: $\sum_{k}\left(i\omega_{m}-\xi_{\mathbf{k}}\right)^{-1}\bigl(i\omega_{m}+\xi_{\mathbf{k}}\bigr)^{-1}\bigl(1+g/\left(i\omega_{m}+\xi_{\mathbf{k}}\right)\bigr)^{-1}$, where $g$ is defined in (73). For small values of the $q$-momentum, the relevant term in the last expression is those which contains the combination:
\begin{eqnarray*}
2m^{-1}\left(\mathbf{k\cdot \widehat{n}}\right)^{2}\left(q^2/2m\right).
\end{eqnarray*}
Then $c\approx T_{c}\sum_{k}\left(i\omega_{m}-\xi_{\mathbf{k}}\right)^{-1}\left(i\omega_{m}+\xi_{\mathbf{k}}\right)^{-3}2m^{-1}k^{2}\cos^{2}{\theta}$. Evaluating the integrals in the $\mathbf{k}$-space, the following result is obtained: $c=7\zeta\left(3\right)N\left(\varepsilon_F\right)\varepsilon_F/12T_{c}^{2}\pi^2$. From Eq. (82), the main contributions in this limit are given in this by: $Q^{\prime}_{\omega}\approx\omega N\left(\varepsilon_F\right)/4\varepsilon_F$, and $Q^{\prime\prime}_{\omega}\approx\omega\pi N\left(\varepsilon_F\right)/8T_{c}$, ($\mu\approx\varepsilon_F$), which define the coefficient $d$.
\newpage
\centerline{\large{\textbf{6. Calculation of the TDGL Coefficients for $T<T_c$}}}
\vspace{0.5cm}
The previous analysis can be extended for the broken symmetry state and high-order expansions in the effective action $S_{eff}\left[\Delta,\bar{\Delta}\right]$.  This formulation permits to make physically motivated approximations for all values of the coupling constants at low temperatures ($T<<T_c$).  
We consider a \emph{uniform} static saddle point $\Delta\left(x\right)=\Delta_0$ which satisfies $\delta S_{eff}/\delta\Delta_0=0$. The relationship between the interaction parameter $g$ and $\Delta_0$ can be obtained by using the Gaussian integration technique. The final result can be expressed as: $1/g=\sum_{\mathbf{k}}\tanh{\left(\beta E_{\mathbf{k}}/2\right)}/2E_{\mathbf{k}}$, where $E_{\mathbf{k}}=\sqrt{\xi_{\mathbf{k}}^2+\Delta_{0}^2}$ (see Eq. 36). The regularization condition for the ultraviolet divergence in the gap equation is established by expressing the final result in terms of the $s$-wave scattering lenght $a_s$, which describes the two-body attraction $g$ in a low density system. The renormalized gap equation then reads (Cfr. Eq. 37):
\begin{equation}
\frac{m}{4\pi a_s}=\sum_{\mathbf{k}}\Bigl[\frac{1}{2\varepsilon_{\mathbf{k}}}-\frac{\tanh{\left(\beta E_{\mathbf{k}}/2\right)}}{2E_{\mathbf{k}}}\Bigr].
\end{equation}
To investigate the fluctuations around the saddle point $\Delta_{0}$, we write $\Delta\left(\mathbf{x},\tau\right)=\Delta_{0}+\eta\left(\mathbf{x},\tau\right)$, and we recall the generic definition for the Green's propagator  in Eq. (63)
\begin{equation}
\mathbf{G^{-1}}=\mathbf{G_{0}^{-1}}\left(\Delta_0\right)+\eta\sigma^{+}+\bar{\eta}\sigma^{-},
\end{equation}
where $\mathbf{G_{0}^{-1}}\left(\Delta_0\right)$ is defined as:
\begin{equation}
\mathbf{G_{0}}\left(\Delta_0\right)=
\begin{pmatrix}
\mathcal{G}_{k}\left(\Delta_0\right)&\mathcal{F}_{k}\left(\Delta_0\right)\\
\mathcal{F}_{k}^{\star}\left(\Delta_0\right)&-\mathcal{G}_{k}^{\star}\left(\Delta_0\right)
\end{pmatrix}
\end{equation}
where $\mathcal{G}_{k}\left(\Delta_0\right)=-\left(i\omega_m+\xi_{\mathbf{k}}\right)/\left(\omega_{m}^2+E_{\mathbf{k}}^2\right)$, and $\mathcal{F}_{k}\left(\Delta_0\right)=\Delta_{0}/\left(\omega_{m}^2+E_{\mathbf{k}}^2\right)$.
The effective action functional must be rewritten in terms of their mean field values $\Delta_0, \bar{\Delta}_0$ and their corresponding fluctuating fields $\eta,\bar{\eta}$. Hence: 
\begin{equation}
S_{eff}
\left[\bar{\eta},\eta\right]=\sum_{q}\Biggl[\frac{\mid\eta_{q}\mid^2}{g}-\mbox{Tr}\ln{\mathbf{G}_{0}^{-1}\left(\Delta_0\right)}+\sum_{n=1}^{\infty}\frac{\left(-1\right)^{n}}{n}\mbox{Tr}\left(\mathbf{G}_{0}\left(\Delta_{0}\right)\mathbf{\Sigma}\right)^{n}\Biggr],
\end{equation}
where $\mathbf{\Sigma}$ is defined as:
\begin{equation}
\mathbf{\Sigma}=
\begin{pmatrix}
0&\eta_{q}\\
\bar{\eta}_{q}&0
\end{pmatrix}.
\end{equation}
The second-order combination in Tr$\left[\mathbf{G}\mathbf{\Sigma}\mathbf{G}^{\prime}\mathbf{\Sigma}^{\prime}\right]$ can be expressed in terms of the \emph{spinors fields} $\eta^{\dag}=\left[\eta^{\star}_q,\eta_{-q}\right]$ and the \emph{correlation propagator} $\widehat{\mathbf{\Pi}}$ through the combination $\eta^{\dag}_{q}\widehat{\mathbf{\Pi}}_{q}\eta_q$, where  $\widehat{\mathbf{\Pi}}$ is defined by
\begin{equation}
\widehat{\mathbf{\Pi}}_{q}\equiv
\begin{pmatrix}
-\sum_{k}\mathcal{G}_{k}\mathcal{G}^{\star}_{k+q}&\sum_{k}\mathcal{F}_{k}\mathcal{F}_{k+q}\\
\sum_{k}\mathcal{F}_{k}^{\star}\mathcal{F}_{k+q}^{\star}&-\sum_{k}\mathcal{G}_{k}^{\star}\mathcal{G}_{k+q}
\end{pmatrix}.
\end{equation}
It is useful to define the spinors compounds like: $\eta_{q}=\left(\lambda_{q}+i\theta_{q}\right)/\sqrt{2}$, with the properties $\lambda_{q}^{\star}=\lambda_{-q}$ and $\theta_{q}^{\star}=\theta_{-q}$, which are usually identified as the phase and amplitude fluctuations, respectively. Under this transformation, the quadratic form is transformed like:
\begin{equation}
\eta^{\dag}_{q}\widehat{\mathbf{\Pi}}_{q}\eta_{q}\equiv
\begin{pmatrix}
\lambda^{\star}_{q}&\theta^{\star}_{q}
\end{pmatrix}
\begin{pmatrix}
\bar{\Pi}^{11}_{q}&i\bar{\Pi}^{12}_{q}\\
-i\bar{\Pi}^{12}_{q}&\bar{\Pi}^{22}_{q}
\end{pmatrix}
\begin{pmatrix}
\lambda_{q}\\ \theta_{q}
\end{pmatrix},
\end{equation}
and
\begin{eqnarray*}
\bar{\Pi}_{q}^{11}=\frac{1}{2}\left(\Pi^{11}_{q}+\Pi^{12}_{q}+\Pi^{21}_{q}+\Pi^{22}_{q}\right), \hspace{0.5cm}\bar{\Pi}_{q}^{12}=\frac{1}{2}\left(\Pi^{11}_{q}-\Pi^{12}_{q}+\Pi^{21}_{q}-\Pi^{22}_{q}\right),
\end{eqnarray*}
\begin{eqnarray*}
\bar{\Pi}_{q}^{22}=\frac{1}{2}\left(\Pi^{11}_{q}-\Pi^{12}_{q}-\Pi^{21}_{q}+\Pi^{22}_{q}\right).
\end{eqnarray*}
The components $\Pi^{ij}_{q}$ are defined in Eq.(95) in terms of the correlations functions $-\sum_{k}\mathcal{G}_{k}\mathcal{G}^{\star}_{k+q}$, $\sum_{k}\mathcal{F}_{k}\mathcal{F}_{k+q}$ and their complex conjugates. Our next task is to obtain approximate expressions (second order in $q$ and in the bosonic frequency $\omega$), for each $\widehat{\Pi}^{ij}_{q}$.
\newpage
\centerline{\emph{\textbf{6.1 Fluctuations in the static Approximation: $i\nu_{n} = 0.$}}}
\vspace{0.5cm}
We identify the following properties in the Green's tensor components at $q_{0}=\nu=0$: $\mathcal{G}^{\star}_{k}=\mathcal{G}_{-k}$, $\mathcal{F}_{k}\mathcal{F}_{k+q}=\mathcal{F}_{k}^{\star}\mathcal{F}^{\star}_{k+q}$ (for $\Delta_{0}=\bar{\Delta}_{0}$). Hence: $\Pi^{11}_{\mathbf{q}}=\Pi^{22}_{\mathbf{q}}$, $\Pi^{12}_{\mathbf{q}}=\Pi^{21}_{\mathbf{q}}$, and $\bar{\Pi}_{\mathbf{q}}^{12}=0$. The quadratic term in the static approximation for the fields $\lambda_{\mathbf{q}}$ and $\theta_\mathbf{q}$ reads explicitly as:
\begin{equation}
\eta^{\dag}_{q}\widehat{\mathbf{\Pi}}_{q}\eta_{q}\mid_{i\nu_{n}=0}\equiv
\begin{pmatrix}
\lambda^{\star}_{\mathbf{q}}&\theta^{\star}_{\mathbf{q}}
\end{pmatrix}
\begin{pmatrix}
\Pi_{\mathbf{q}}^{11}+\Pi_{\mathbf{q}}^{12}&0\\
0&\Pi_{\mathbf{q}}^{11}-\Pi_{\mathbf{q}}^{12}
\end{pmatrix}
\begin{pmatrix}
\lambda_{\mathbf{q}}\\ \theta_{\mathbf{q}}
\end{pmatrix}.
\end{equation}
The coefficients in the diagonal form in Eq. (97) can be reduced as:
\begin{eqnarray*}
\Pi_{\mathbf{q}}^{11}\pm\Pi_{\mathbf{q}}^{12}=
\end{eqnarray*}
\begin{eqnarray}
\sum_{\mathbf{k}}\frac{E_{\mathbf{k+q}}\left(\xi_{\mathbf{k}}\xi_{\mathbf{k+q}}-E_{\mathbf{k}}^2\mp\Delta^{2}_{0}\right)\tanh{\left(\beta E_{\mathbf{k}}/2\right)}-E_{\mathbf{k}}\left(\xi_{\mathbf{k}}\xi_{\mathbf{k+q}}-E_{\mathbf{k+q}}^2\mp\Delta^{2}_{0}\right)\tanh{\left(\beta E_{\mathbf{k+q}}/2\right)}}{2E_{\mathbf{k}}E_{\mathbf{k+q}}\left(E_{\mathbf{k}}^2-E_{\mathbf{k+q}}^2\right)}.
\end{eqnarray}
For small values of the  bosonic momentum $\mathbf{q}$, ($q^{2}/m<<\mbox{min}_{\mathbf{k}}\lbrace E_{\mathbf{k}}\rbrace$) the last expression becomes: $\Pi_{\mathbf{q}}^{11}\pm\Pi_{\mathbf{q}}^{12}\approx\bar{c}_{\pm}\left(\Delta_0,T,\mu\right)\left(q^{2}/2m\right)$, where $\bar{c}_{\pm}$ is formally defined in Appendix A. The particular case in which $T=T_c$, ($\Delta_0=0$) the last results reproduce the expressions for $a$ and $c$ in Eqs. (71) and (75) respectively. In the opposite limit (i.e. $T=0$), and for $\mathbf{q}=0$, the set of Equations (98) are reduced to: $\Pi_{0}^{11}-\Pi_{0}^{12}=0$, which ensures that the phase mode for $\mathbf{q}$ is gapless, i.e., the Goldstone mode, and $\Pi_{0}^{11}+\Pi_{0}^{12}=\sum_{\mathbf{k}}\Delta_{0}^{2}/2E_{\mathbf{k}}^3$, which correspond to the results obtained by Engelbrecht et-al in Ref. [10]. \\
The effective action Eq. (97) may be scripted as:
\begin{equation}
S_{eff}\left[\bar{\eta},\eta\right]=S_{0}+\sum_{\mathbf{q}}\eta^{\dag}_{\mathbf{q}}\left(K^{-1}_{\mathbf{q}}\right)\eta_{\mathbf{q}},
\end{equation}
where the matrix $K^{-1}_\mathbf{q}=\frac{1}{2}\left(g^{-1}+\widehat{\mathbf{\Pi}}_{\mathbf{q}}\right)$ is diagonal in this approximation, and $S_0=-\sum_{q}\mbox{Tr}\ln{\mathbf{G}_{0}^{-1}\left(\Delta_0\right)}$. Let us define the generating functional $\mathcal{Z}$ for the bilinear form as: 
\begin{equation}
\mathcal{Z}=\mathcal{Z}_{0}\int\mathcal{D}\left[\bar{\eta},\eta\right]\exp{\left[-\sum_{\mathbf{q}}\eta^{\dag}_{\mathbf{q}}K^{-1}_{\mathbf{q}}\eta_{\mathbf{q}}\right]}=\mathcal{Z}_{0}\prod_{\mathbf{q}}\det{\left(K_\mathbf{q}\right)}.
\end{equation}
where $\mathcal{Z}_0=\exp{\left(-S_0\right)}$, and $S_0$ is defined in the textline after Eq. (99). Quantities of physical interest such as the \emph{free energy} can now be directly computed by using $\mathcal{F}=-T\ln{\mathcal{Z}}$\cite{Feym}. Inserting the compounds of $K_{\mathbf{q}}$ into the second term in the r.h.s of (100), we get:
\begin{equation}
\mathcal{F}\left(\Delta_0,\mu,T\right)=\mathcal{F}_{0}\left(\Delta_0,\mu,T\right)+\frac{1}{\beta}\sum_{\mathbf{q}}\ln{\Biggl[\frac{\left(g^{-1}+\Pi^{11}_{\mathbf{q}}\right)^2-\left(\Pi^{12}_{\mathbf{q}}\right)^2}{4}\Biggr]},
\end{equation}
where $\mathcal{F}_0=\beta^{-1}S_{0}$, while the relations for $\Pi^{ij}_{\mathbf{q}}$ are explicitly encoded in Eq. (98) and in Appendix A. In order to obtain the quadratic average fluctuations in the field $\eta$, we consider the path integral representation in terms of the amplitude and phase components as follows: (notice that $S_0$ does not depend on the elements $K_{ij}^{-1}$)
\begin{equation}
\ln{\mathcal{Z}}=-S_{0}+\sum_{\mathbf{q}}\ln{\mathcal{Z}_{\mathbf{q}}}, \hspace{1cm}\mathcal{Z}_{\mathbf{q}}\equiv\mathcal{Z}^{\lambda}_{\mathbf{q}}\mathcal{Z}^{\theta}_{\mathbf{q}} 
\end{equation}
where the functions $\mathcal{Z}^{\lambda}$, $\mathcal{Z}^{\theta}$ are respectively given by: 
\begin{eqnarray*}
\mathcal{Z}^{\lambda}_{\mathbf{q}}=\int\mathcal{D}\left[\lambda^{\star}_{\mathbf{q}},\lambda_{\mathbf{q}}\right]\mbox{\large{e}}^{-K_{11}\left(\mathbf{q}\right)^{-1}\mid\lambda_{\mathbf{q}}\mid^2}, \hspace{1cm} \mathcal{Z}^{\theta}_{\mathbf{q}}=\int\mathcal{D}\left[\theta^{\star}_{\mathbf{q}},\theta_{\mathbf{q}}\right]\mbox{\large{e}}^{-K_{22}\left(\mathbf{q}\right)^{-1}\mid\theta_{\mathbf{q}}\mid^2}. 
\end{eqnarray*}
Therefore, there is straightforward to show that:
\begin{equation}
\langle\mid\eta\mid^2\rangle=-\sum_{\mathbf{q}}\left(\frac{\delta\ln{\mathcal{Z}^{\lambda}}}{\delta\left(K_{11}^{-1}\right)}+\frac{\delta\ln{\mathcal{Z}^{\theta}}}{\delta\left(K_{22}^{-1}\right)}\right),
\end{equation}
or
\begin{equation}
\langle\mid\eta\mid^2\rangle_{\omega=0}=4\sum_{\mathbf{q}}\frac{g^{-1}+\Pi^{11}_{\mathbf{q}}}{\left(g^{-1}+\Pi^{11}_{\mathbf{q}}\right)^2-\left(\Pi^{12}_{\mathbf{q}}\right)^2}.
\end{equation}
The numerical evaluation of this quantity as a function of the temperature will be discussed elsewhere. An improved approximation for the order parameter is defined by adding the fluctuating term $\eta$ obtained in Eq. (104): $\Delta=\Delta_0+\sqrt{\langle\mid\eta\mid^2\rangle}$. This correction can be interpreted as a \lq\lq pseudogap" the context of simple MFA theories.
\newpage
\centerline{\emph{\textbf{6.2 Dynamics of Fluctuations: A Generic expression}}}
\vspace{0.5cm}
For the case in which the set of bosonic frequencies $i\nu_{n}$ is different of zero, the dynamical approximation is taken into consideration directly from the formula (99). The tensor $K_{q}^{-1}$ might be separated into its \lq\lq static" component, which has been collected into $K_{\mathbf{q}}^{-1}$ (See the textline right after Eq. 99), and its frequency-dependent part denoted as $K_{\omega}^{-1}$, where $\omega$ is the real frequency that appears after analytic continuation on $i\nu$, under $i\nu\rightarrow\omega+i0^+$. Hence, $K_{q}^{-1}=K_{\mathbf{q}}^{-1}+K_{\omega}^{-1}$, with
\begin{equation}
\left(K^{11}_{\omega}\right)^{-1}=-\sum_{\mathbf{k}}\frac{\left(1-\Delta_{0}^{2}/E_{\mathbf{k}}^2\right)\tanh{\left(\beta E_{\mathbf{k}}/2\right)}}{2E_{\mathbf{k}}\left(1-\omega^2/4E^{2}_{\mathbf{k}}\right)},
\end{equation}
\begin{equation}
\left(K^{22}_{\omega}\right)^{-1}=-\sum_{\mathbf{k}}\frac{\tanh{\left(\beta E_{\mathbf{k}}/2\right)}}{2E_{\mathbf{k}}\left(1-\omega^2/4E^{2}_{\mathbf{k}}\right)},
\end{equation}
\begin{equation}
\left(K^{12}_{\omega}\right)^{-1}=i\sum_{\mathbf{k}}\frac{\omega\xi_{\mathbf{k}}\tanh{\left(\beta E_{\mathbf{k}}/2\right)}}{4E^{3}_{\mathbf{k}}\left(1-\omega^2/4E^{2}_{\mathbf{k}}\right)},
\end{equation}
with $\left(K^{21}_{\omega}\right)^{-1}=-\left(K^{12}_{\omega}\right)^{-1}$. The expression (100) for the partition function $\mathcal{Z}$ might be modified as:
\begin{equation}
\ln{\mathcal{Z}}=-S_{0}+\sum_{\mathbf{q}}\int d\omega\ln{\left[\det\left(K_{q}\right)\right]},
\end{equation}
where
\begin{eqnarray*}
\det{\left(K_{\mathbf{q},\omega}^{-1}\right)}=
\end{eqnarray*}
\begin{equation}
\frac{1}{4}\left[g^{-1}+\Pi^{11}_{\mathbf{q}}+\Pi^{12}_{\mathbf{q}}+\left(K^{11}_{\omega}\right)^{-1}\right]\left[g^{-1}+\Pi^{11}_{\mathbf{q}}-\Pi^{12}_{\mathbf{q}}+\left(K^{22}_{\omega}\right)^{-1}\right]+\frac{1}{4}\left[\left(K^{12}_{\omega}\right)^{-1}\right]^2.
\end{equation}
The last expression provides the core result in this report: The complete set of self-consistent equations that describes the Helmhontz free energy associated to Gaussian fluctuations in the field $\eta$, via the relationship $\mathcal{F}\left(\Delta_0,T,\mu\right)=-\beta^{-1}\ln{\mathcal{Z}}$, for any temperature between $T$ and $T_c$, and under the regularization condition in the BSC-BEC crossover scenario (Eq. 90). A similar expression to Eq. (104) is obtained for the full dynamics in the fluctuating field $\eta\left(\omega,\mathbf{q}\right)$:
\begin{equation}
\langle\mid\eta\mid^{2}\rangle_{\mathbf{q},\omega}=\frac{\mbox{Tr}\lbrace K_{\mathbf{q},\omega}^{-1}\rbrace}{\det{\left(K_{\mathbf{q},\omega}^{-1}\right)}},
\end{equation}
where Tr$\lbrace{\cdot\rbrace}$ indicates the Trace over the embraced quantity. 
\newpage
\centerline{\large{\textbf{Concluding Remarks}}}
\vspace{0.5cm}
In this paper we have shown how to formulate and derivate the quadratic Gaussian fluctuations associated to the perturbations in the order paramerter in a uniform Fermi gas, for \emph{any} temperature below $T_c$, by using path integral formulation. The average estimation for these fluctuations was obtained in \emph{exact} form for the static case, while the partition function was drafted in a general way in terms of the BCS correlation propagators for the dynamical approach.\\
The possibilities for a further work are listed in the following terms:
\begin{enumerate}
\item[$\triangleright$] Numerical analysis for the final expression for $\langle\mid\eta\mid^2\rangle$, as a function of the temperature and/or different values of the chemical potential $\mu$, and the collective excitations associated to the phase and amplitude modes in the BCS-BEC crossover region.
\item[$\triangleright$] Perturbative analysis up to order four ($\mathcal{O}\left(\mid\eta\mid^4\right)$) in the effective action functional $S_{eff}$.
\item[$\triangleright$] TDGL treatment for trapped Fermi gas.
\item[$\triangleright$] A long-term project might certainly be orientated in obtaining analogous expressions for a Fermi gas in the Feshbach resonance scenario \cite{Timmermans}. 
\end{enumerate}
\emph{\textbf{Acknowledgements}}
\\
I am very grateful to Prof. A. Griffin (\emph{in memoriam}) for his recommendation of this topic and fruitful discussions. This work was partially supported by NSERC Canada.
\newpage
\centerline{\LARGE{Appendix A:}}
\centerline{\LARGE{Definition of the \lq\lq kinetic" coefficient $\bar{c}_{\pm}$ in Eq. (98).}}
\vspace{0.5cm}
The expansion in the second order in the wavector $\mathbf{q}$ leads to the expressions \footnote{ There is useful to consider the following results: ($E=E_{\mathbf{k}},\xi=\xi_{\mathbf{k}}, E^{\prime}=E_{\mathbf{k+q}}, \xi^{\prime}=\xi_{\mathbf{k+q}}$)
\begin{eqnarray*}
E^{\prime 2}-E^{2}\approx 4x\left(1+x\right)E^2, \hspace{0.25cm} x=Q/4E^2-Q^{2}/16E^4, \hspace{0.25cm} Q=2h^{\prime}\xi+h^{\prime 2},
\hspace{0.25cm}h^{\prime}=\xi^{\prime}-\xi, \hspace{0.25cm}h^{\prime}=\left(q^2+2q\left(\mathbf{k\cdot\widehat{n}}\right)\right)/2m.
\end{eqnarray*}
}:
\begin{eqnarray*}
\bar{c}_{\pm}\left(\Delta_0,T,\mu\right)=\sum_{\mathbf{k}}\Biggl[\left(\frac{\bar{X}_{\mathbf{k}}}{4E_{\mathbf{k}}^2}-\frac{\beta\bar{Y}_{\mathbf{k}}}{8E_{\mathbf{k}}}\right)Z^{\pm}_{\mathbf{k}}+\frac{\beta^{2}\bar{X}_{\mathbf{k}}\bar{Y}_{\mathbf{k}}}{2}Z_{\mathbf{k}}^{\prime\pm}\Biggr],
\end{eqnarray*}
with
\begin{eqnarray*}
Z^{+}_{\mathbf{k}}=\frac{\xi_{\mathbf{k}}}{E_{\mathbf{k}}}\left(1-3\frac{\Delta_{0}^2}{E_{\mathbf{k}}^2}\right)+\frac{\left(\mathbf{k\cdot\widehat{n}}\right)^2}{mE_{\mathbf{k}}}\left(-1+10\frac{\Delta_{0}^2}{E^{2}_{\mathbf{k}}}\left(1-\frac{\Delta_{0}^2}{E^{2}_{\mathbf{k}}}\right)\right),
\end{eqnarray*}
\begin{eqnarray*}
Z_{\mathbf{k}}^{-}=\frac{\xi_{\mathbf{k}}}{E_{\mathbf{k}}}+\frac{\left(\mathbf{k\cdot\widehat{n}}\right)^2}{mE_{\mathbf{k}}}\left(-1+3\frac{\Delta_{0}^2}{E^{2}_{\mathbf{k}}}\right),
\end{eqnarray*}
\begin{eqnarray*}
Z_{\mathbf{k}}^{\prime+}=\frac{\Delta_{0}^{2}\xi_{\mathbf{k}}}{4E^{3}_{\mathbf{k}}}+\frac{\left(\mathbf{k\cdot\widehat{n}}\right)^2}{4mE_{\mathbf{k}}}\left(1-4\frac{\Delta_{0}^2}{E_{\mathbf{k}}^2}+6\frac{\Delta_{0}^4}{E_{\mathbf{k}}^4}\right),
\end{eqnarray*}
\begin{eqnarray*}
Z_{\mathbf{k}}^{\prime-}=\frac{\left(\mathbf{k\cdot\widehat{n}}\right)^2}{4mE_{\mathbf{k}}}\left(1-\frac{\Delta_{0}^2}{E_{\mathbf{k}}^2}+2\frac{\Delta_{0}^4}{E_{\mathbf{k}}^4}\right),
\end{eqnarray*}
and $\bar{X}=\tanh{\left(\beta E/2\right)}$, $\bar{Y}=\mbox{sech}^{2}\left(\beta E/2\right)$. Notice that for $\Delta_{0}=0$, $Z^{+}=Z^{-}, Z^{\prime +}=Z^{\prime -}$ and this results will reduce to those obtained by Randeira et-al. [8].
\newpage
\centerline{\LARGE{Appendix B:}}
\centerline{\LARGE{Collective Modes}}
\vspace{0.5cm}
By using the results in appendix A, there is possible to obtain the relationship for the collective excitations $\omega\left(\mathbf{q}\right)$ by solving $\det{\left(K^{-1}_{\mathbf{q},\omega}\right)}=0$ in Eq. (109). It is clear that for small values of momentum and frequency, 
\begin{eqnarray*}
\left(K_{q}^{11}\right)^{-1}\approx A-D\omega^{2}+\bar{c}_{+}\left(\frac{q^2}{2m}\right), \hspace{0.5cm}\left(K_{q}^{22}\right)^{-1}\approx-R\omega^2+\bar{c}_{-}\left(\frac{q^2}{2m}\right), \hspace{0.5cm} \left(K_{q}^{12}\right)^{-1}\approx iB\omega.
\end{eqnarray*}
with $A=\sum_{\mathbf{k}}\Delta_{0}^{2}\bar{X}/2E^3$, $D=\sum_{\mathbf{k}}\left(1-\Delta_{0}^{2}/E^2\right)\bar{X}/8E^{3}$, $R=\sum_{\mathbf{k}}\bar{X}/8E^3$, and $B=\sum_{\mathbf{k}}\xi\bar{X}/4E^{3}$, with reduced notation in the coefficients $A,B,D,R$, and $c_{\pm}, \bar{X}$ defined in Appendix A. Therefore:
\begin{eqnarray*}
\left[A-D\omega^2+\bar{c}_{+}\left(\frac{q^2}{2m}\right)\right]\left[-R\omega^2+\bar{c}_{-}\left(\frac{q^2}{2m}\right)\right]-B^2\omega^2=0.
\end{eqnarray*}
Neglecting terms of the form $\omega^4$, $q^4$ and $\omega^{2}q^2$, we get a first approach for the collective excitations modes: $\omega\left(\mathbf{q}\right)=v_{s}\left(T,\Delta_{0},\mu\right)\mid\mathbf{q}\mid$, with the  \lq\lq sound velocity" given by:
\begin{eqnarray*}
v_{s}\left(T,\Delta_{0},\mu\right)=\sqrt{\frac{A\bar{c}_{-}}{2m\left(B^{2}+AR\right)}}.
\end{eqnarray*}
For arbitrary coupling, the integrals involved in $A,B$, etc, have to be evaluated numerically. Integrals involving in $A, B, \bar{c}_{\pm}...$, peak around the Fermi surface at $T\rightarrow 0$, for the weak-coupling limit. Specifically, $A=0$, $D=R=7N\left(\varepsilon_{F}\right)\zeta\left(3\right)/16\pi^{2}T_{C}^2$, $c_{+}=c_{-}=4\varepsilon_{F}D/3$ and $B=\left(2\sqrt{2}-1\right)\zeta\left(3/2\right)N\left(\varepsilon_{F}\right)/\left(16\pi T_{C}\varepsilon_{F}\right)^{1/2}$. 
Under this conditions, the dispersion relationship for $\omega_{q}$ takes the linear form $\omega_{q}=v_{s}q$, and $v_s$, the \textit{sound} velocity in the condensate approaches to the value $v_{s}=v_{F}/\sqrt{3}$, ($v_{F}=\left(2\varepsilon_{F}/m\right)^{1/2}$ as the Fermi velocity), which corresponds to the well-known result for the \textit{phase} mode.

\end{document}